# Temperature Dependence of Rashba Spin-orbit Coupling in Quantum Wells

P.S. Eldridge[1], W.J.H. Leyland[2], P.G. Lagoudakis[1], O.Z.Karimov[1], M. Henini[3], D.Taylor[3], R.T. Phillips[2] and R.T. Harley[1]

[1]School of Physics and Astronomy, University of Southampton, Southampton, SO17 1BJ, UK.
[2]Cavendish Laboratory, Madingley Road, Cambridge, CB3 0HE, UK.
[3]School of Physics and Astronomy, University of Nottingham, Nottingham, NG7 4RD, UK.

**Abstract**

We perform an all-optical spin-dynamic measurement of the Rashba spin-orbit interaction in (110)-oriented GaAs/AlGaAs quantum wells. The crystallographic direction of quantum confinement allows us to disentangle the contributions to spin-orbit coupling from the structural inversion asymmetry (Rashba term) and the bulk inversion asymmetry. We observe an unexpected temperature dependence of the Rashba spin-orbit interaction strength that signifies the importance of the usually neglected higher-order terms of the Rashba coupling.

Intense worldwide interest is being focused on new semiconductor spintronic and spin-optronic quantum devices in which electronic spin replaces charge for data processing or is used to control optical polarisation. Indeed manipulation of electron spins has been signposted as the preferred route to quantum computing [1]. Progress towards realistic devices depends on engineering the spin-orbit interactions that result in effective magnetic fields seen by the electron spins under application of external electric fields. A classic example is the Datta and Das spin transistor, wherein the precession of spin polarised carriers confined in a plane is controlled by a gate voltage which tunes the Rashba or structural inversion asymmetry (SIA) component of spin-orbit coupling [2]. Whereas the ability to tune the Rashba (SIA) coupling strength originates in the field-induced spin-splitting of the electronic bands, other contributions to the spin-splitting, from bulk inversion asymmetry (BIA or Dresselhaus coupling) and natural interface asymmetry in heterostructures (NIA), complicate direct characterisation of the Rashba term. Disentangling and evaluating the contribution of the different spin-orbit coupling mechanisms to the spin-splitting of the electronic bands is of utmost importance for the engineering of spintronic devices. Most of the studies that focus on the characterisation of the Rashba coupling have either neglected the contribution of the Dresselhaus coupling or have measured the ratio of the strength of the two mechanisms [3,4]. Here we have designed and grown quantum well heterostructures that utilise crystal asymmetries in a way that allows us, through combined optical measurements of spin relaxation and of electron mobility, to separate the terms and measure directly the strength of the Rashba coupling. The values are in good quantitative agreement with k.p-theoretical estimations but unexpectedly, we observe a temperature dependence of the Rashba coefficient that signifies the importance of usually neglected higher order terms in the Rashba spin-orbit coupling [5].

The spin relaxation of a non-equilibrium population of electron spins in non-centrosymmetric semiconductors may involve several mechanisms [1], the dominant one in all except $p$-type material is that identified by D'yakonov, Perel' and Kachorovskii (DPK)[6,7]. The driving force for spin reorientation and therefore loss of spin memory is the combination of inversion asymmetry and the spin-orbit interaction; as an electron propagates its spin tends to precess. The corresponding precession vector $\Omega(\mathbf{k})$, which describes conduction band spin-splitting, varies in magnitude and direction according to the electron's wavevector $\mathbf{k}$. Strong scattering of the electron wavevector randomises the precession and causes spin relaxation. The vector $\Omega(\mathbf{k})$ is the sum of the three components described above and denoted $\Omega^{BIA}(\mathbf{k})$, $\Omega^{SIA}(\mathbf{k})$ and $\Omega^{NIA}(\mathbf{k})$ [1,8]. For a quantum well grown on a (110)-oriented substrate the interface component $\Omega^{NIA}(\mathbf{k})$ is zero [1,9]. Furthermore, since electron motion is confined to the (110) plane, $\Omega^{BIA}(\mathbf{k})$ is, by symmetry, normal to the plane, parallel or antiparallel to the growth axis [110], for all electron wavevectors. Therefore, the *BIA* term makes no contribution to spin relaxation along the growth axis [7]. Thus using this configuration we can measure directly the spin relaxation due to the structural inversion asymmetry alone and derive the strength of the Rashba coupling, $\Omega^{SIA}(\mathbf{k})$.

The DPK mechanism gives the relaxation rate for the component of a spin population along a particular axis, $i$, as [1]

$$\tau_{s,i}^{-1} = <\Omega_\perp^2> \tau_p^*, \qquad (<|\Omega|> \tau_p^* << 1), \qquad (1)$$

where $<\Omega_\perp^2>$ is the averaged square component of $\Omega(\mathbf{k})$ *perpendicular* to the axis $i$ taken over the spin-oriented population and $\tau_p^*$ is the momentum scattering time of an electron. Furthermore for a symmetrical quantum well, to first order, the *SIA* term has the form [8,10]

$$\Omega^{SIA}(\mathbf{k}) = \alpha(e/\hbar)\, \mathbf{F} \times \mathbf{k} \qquad (2)$$

where $\alpha$ is the Rashba coefficient and **F** the applied electric field. By applying **F** along the growth axis, $\Omega^{SIA}(\mathbf{k})$ lies in the quantum well plane and thus causes DPK spin relaxation along the growth axis. Squaring eq. 2 and taking the thermal average assuming $k_B T >> \hbar \Omega^{SIA}$ gives

$$<(\Omega^{SIA})^2> = \alpha^2 \frac{e^2}{\hbar^2} F^2 <k^2> = \alpha^2 \frac{e^2}{\hbar^2} F^2 \frac{2m^* k_B T}{\hbar^2} \qquad (3)$$

where $m^*$ is the electron effective mass and $k_B$ is Boltzmann's constant. Substitution eq.3 in eq. 1 shows that the spin relaxation rate should be linear in $F^2$ and the Rashba coefficient becomes

$$\alpha = \frac{\hbar^2}{eF} (2m^* k_B T \tau_s \tau_p^*)^{-\frac{1}{2}} \qquad (4)$$

Thus we can obtain the Rashba coefficient $\alpha$ from combined measurements of momentum relaxation time and of spin relaxation in applied electric field.

Our sample consists of a GaAs/Al$_{0.4}$Ga$_{0.6}$As $p^+$-$i$-$n^+$ structure grown on a semi-insulating (110)-oriented GaAs substrate. The insulating portion of the structure comprises 100nm layers of undoped AlGaAs on each side of a stack of 20 7.5 nm undoped GaAs quantum wells with 12 nm undoped AlGaAs barriers. For the pump-probe measurements of the spin-relaxation time, $\tau_s$, a portion of the wafer is processed into a mesa device 400 microns in diameter with an annular metal contact to the top $n^+$ layer to allow optical access to the quantum wells and a second contact to the lower $p^+$ layer. The electric field is varied by means of applied bias voltage. The total applied electric field **F** is obtained as the sum of that due to the bias and that to the built-in electric field of the *pin* structure which we calculate to be $2.80 \; 10^6$ Vm$^{-1}$. All measurements are made at temperatures above 80K in order to reveal effects of free electrons rather than excitons in the quantum wells [11]. Photocurrent measurements in the *pin* device as a function of reverse bias and of photon energy reveal evidence of resonant tunnelling between n=1 and n=2 confined electron states in adjacent wells above about 3 volts. In the measurements reported here we therefore concentrate on applied bias less than 3 volts equivalent to $\mathbf{F} \leq 8 \; 10^6$ Vm$^{-1}$. Under these conditions the electrons are resonantly excited into the n=1 confined state of a quantum well and can be considered to remain there until recombination, thermal excitation over the barriers being negligible. The momentum relaxation time, $\tau_p^*$, is obtained from measurement of the electron diffusion coefficient using a transient spin-grating method on an unprocessed portion of the same wafer. This gives $\tau_p^*$ at only one value of transverse field namely the built-in field, $2.80 \; 10^6$ Vm$^{-1}$, and in our analysis we assume that it has no significant dependence on the field.

The spin relaxation of the electrons is investigated using a picosecond-resolution polarised pump-probe reflection technique (see figure.1a) [12]. Wavelength-degenerate circularly polarised pump and delayed linearly polarised probe pulses from a mode-locked Ti-sapphire laser are focused at close to normal incidence on the sample and tuned to the n=1 heavy-hole to conduction band transition. The pulse duration is ~1.5 ps and repetition frequency is 75 MHz. Absorption of each pump pulse generates a photoexcited population of electrons spin-polarised along the growth axis. The time evolution of the photoexcited population and of the spin polarisation $<S_z>(t)$ are monitored by measuring pump-induced changes of, respectively, probe reflection $\Delta R$ and of probe polarisation rotation $\Delta\theta$ as functions of probe pulse delay. These measurements are combined to give the spin relaxation rate $\tau_s^{-1}$. They also show that the recombination time of photoexcited carriers, $\tau_r$, is typically five times longer than $\tau_s$. The pump beam intensity is typically 0.5 mW focused to a 60-micron-diameter spot giving an estimated photoexcited spin-polarised electron density $N_{ex} \sim 10^9$ cm$^{-2}$; the probe intensity is 25% of the pump. Figure 2 shows $\tau_s^{-1}$ vs $\mathbf{F}^2$ for three different temperatures. For fields above $\sim 3 \; 10^6$ V m$^{-1}$ the relationship is linear within experimental uncertainty; the lines represent best fits to the experimental points from which the Rashba coefficient is obtained. For lower values of field, $\tau_s^{-1}$ tends to a constant value which may be associated with imperfections of the interfaces [13] or Bir-Aronov-Pikus spin relaxation [1] due to accumulation of photoexcited carriers under forward bias of the *pin* structure.

The spin-grating measurements (see figure 1b) [14, 15] are made using twin 0.5 mW pump beams from a 200 fs pulse-length mode-locked Ti-sapphire laser tuned to the n=1 valence-conduction band transition. The beams are linearly polarised at 90 degrees to one another and incident on the sample at ±4.1 degrees to the normal. This produces interference fringes of polarisation but not intensity, resulting in a transient grating of spin population with a pitch $\Lambda \sim$ 5.7 microns. The focal spot size on the sample is again of order 60 microns giving excitation density $N_{ex} \sim 10^9$ cm$^{-2}$. The decay rate of the amplitude of such a grating is given by [14]

$$\Gamma = D_s \frac{4\pi^2}{\Lambda^2} + \tau_s^{-1} + \tau_r^{-1} \qquad (5)$$

where $D_s$ is the electron spin-diffusion coefficient. The decay is monitored by measuring first-order diffraction, in reflection geometry, using a delayed 0.25 mW linear polarised probe beam from the same laser, incident on the sample at normal incidence. Signal to noise is enhanced by use of an optical heterodyne detection scheme [15]; the decay rate of the diffracted intensity is $2\Gamma$. Since the sample is undoped we can equate the electron spin-diffusion coefficient to the diffusion coefficient $D_e$ and obtain the electron mobility from the Einstein equation $\mu = (e/k_BT)D_e$. Figure 3 shows an example of a measured decay together with the extracted values of electron mobility as a function of temperature. The grating decay rate (and therefore $D_e$) is found to be insensitive to temperature so that $\mu \sim T^{-1}$ (see Fig.3). This temperature dependence is as expected for a non-degenerate two-dimensional electron system, that is with constant density of states, and dominant phonon scattering with probability $\sim k_BT$. From the mobility we obtain the ensemble momentum relaxation time $\tau_p = m^*\mu/e$ and since we are dealing with intrinsic material with negligible electron-electron scattering we can equate this to the momentum scattering time $\tau_p^*$ [1,16]

Figure 4 shows the values of the Rashba coefficient obtained by combining the two sets of measurements. The value is approximately 0.1 nm$^2$ however there is a clear upward trend which is consistent with a linear increase of the Rashba coefficient with electron kinetic energy. The solid curve is based on the 8-band k.p treatment given by Winkler [10]. The spin splitting is given as

$$|\Omega^{SIA}| = \alpha \frac{e}{\hbar}|k|F = \frac{e^2 P^2}{3\hbar^2}\left(\frac{1}{E_g^2} - \frac{1}{(E_g+\Delta_0)^2}\right)|k|F_v \qquad (6)$$

where $E_g$ is the band gap of GaAs, $\Delta_0$ the spin-orbit splitting in the valence band and $P$ is Kane's momentum matrix element [17]. $F_v$ is the effective electric field in the *valence band* and is given by [10]

$$F_v = F\left(1 + \frac{\Sigma_v}{\Sigma_c}\right) \approx 1.67F \qquad (7)$$

$\Sigma_v$ and $\Sigma_c$ being the valence and conduction band offsets between GaAs and AlGaAs which we take to be in the ratio 3: 2. The weak temperature dependence of the curve results from the temperature dependence of $E_g$ and recently identified temperature dependence of $P$ [18]. This theoretical estimate is in satisfactory agreement with the magnitude of the experimental values of Rashba coefficient but it does not reproduce the observed temperature dependence. We note that extension of the k.p treatment to 14 bands, following [10,19], increases the calculated values by less than 2%.   The dotted curve in Fig. 4 is the k.p theory plus an empirical term linear in temperature. Within the experimental uncertainties this reproduces the data reasonably well. As the photoexcited electron gas is nondegenerate, the additional linear temperature dependence suggests that the Rashba coefficient has an additional approximately linear dependence on the electrons' kinetic energy. This is reminiscent of the dependence of the Zeeman splitting on kinetic energy [20] which in turn gives rise to the observed increase of the effective electron g-factor with quantum confinement in GaAs/AlGaAs quantum wells [21]. The observed dependence signifies the importance of the usually neglected higher order terms in the Rashba Hamiltonian.

In conclusion, by combined measurements of spin relaxation and of electron mobility in an undoped and nominally inversion symmetric (110)-oriented quantum wells in a *pin* structure, we have been able to investigate directly the electric field spin-splitting of the conduction band without interfering effects from bulk inversion asymmetry. The observed splittings are in qualitative agreement with a theoretical k.p calculation [10] but also reveal an unexpected significant temperature dependence. The observation of the temperature dependence of the Rashba coefficient is important for developing an understanding of the fundamental interactions in semiconductor nanostructures and for engineering spintronic devices.

We acknowledge useful discussions with Roland Winkler and Xavier Cartoixa and financial support of the Engineering and Physical Sciences Research Council (EPSRC).

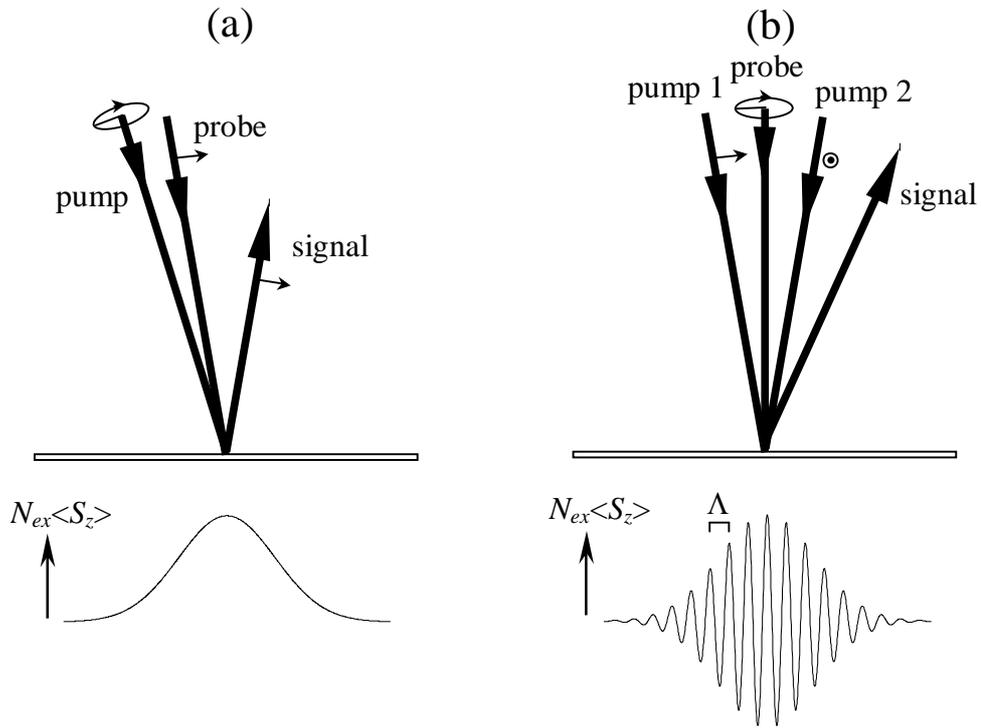

**Figure 1.** Experimental configurations for measurements of (a) spin relaxation and (b) spin diffusion. Upper diagrams, incident pump and probe beams and their polarisations; lower diagrams, profile of focussed pump spot with (a) a spin polarised population and (b) spin grating with spacing $\Lambda$.

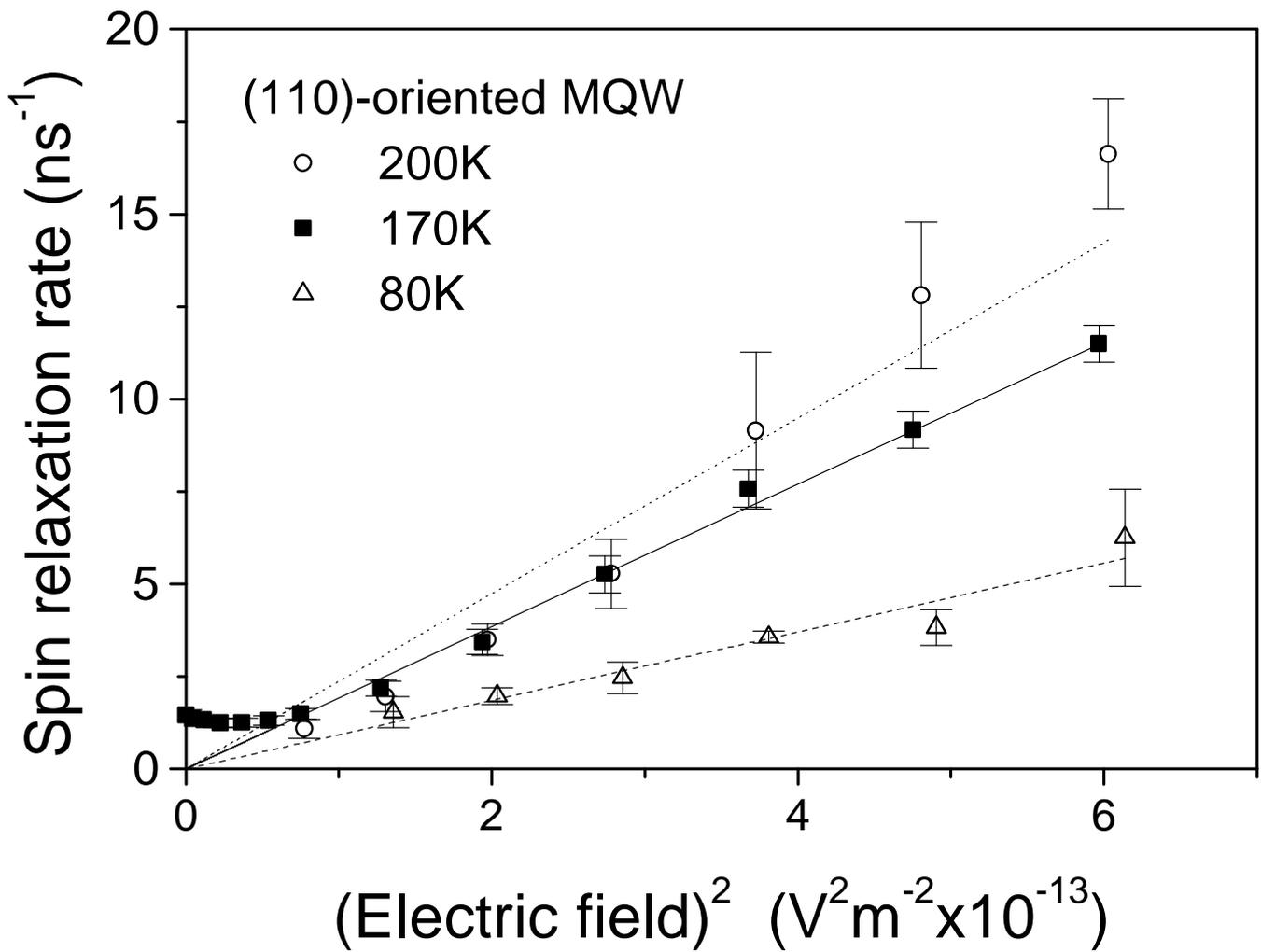

**Figure 2.** Electric field dependence of spin relaxation rate at three temperatures. The high field linear regions of the graphs extrapolate to the origin and the slopes are used to determine the Rashba coefficient

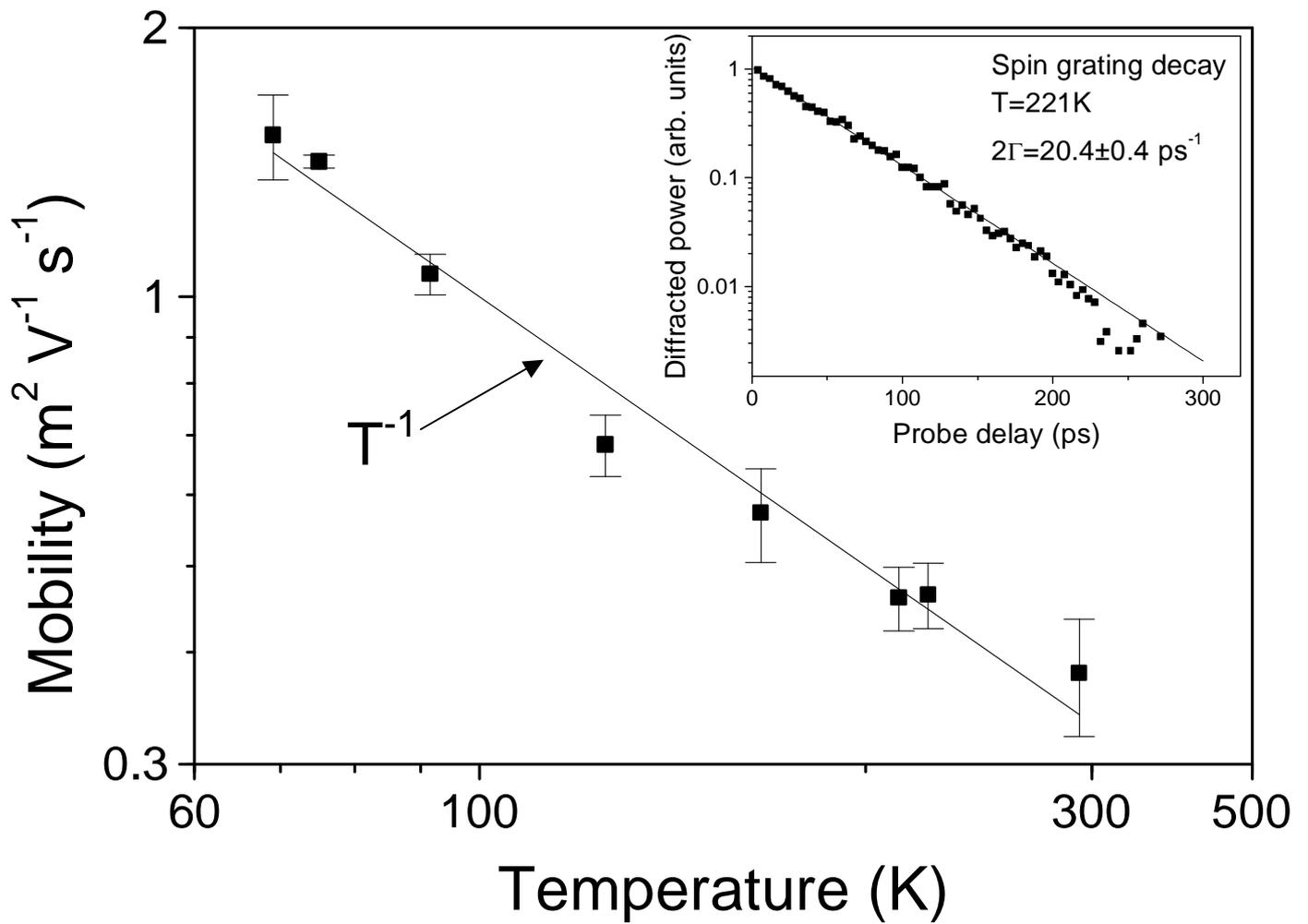

**Figure 3.** Logarithmic plot of the electron mobility, in a sample from the same wafer as data of Fig.2, determined by the spin-grating method. The electric field is the built-in field of the *pin* structure, $2.80 \; 10^6$ Vm$^{-1}$. The $T^{-1}$ temperature dependence is as expected for a non-degenerate two-dimensional electron system with dominant phonon scattering. Inset is a typical grating decay signal at 221 K.

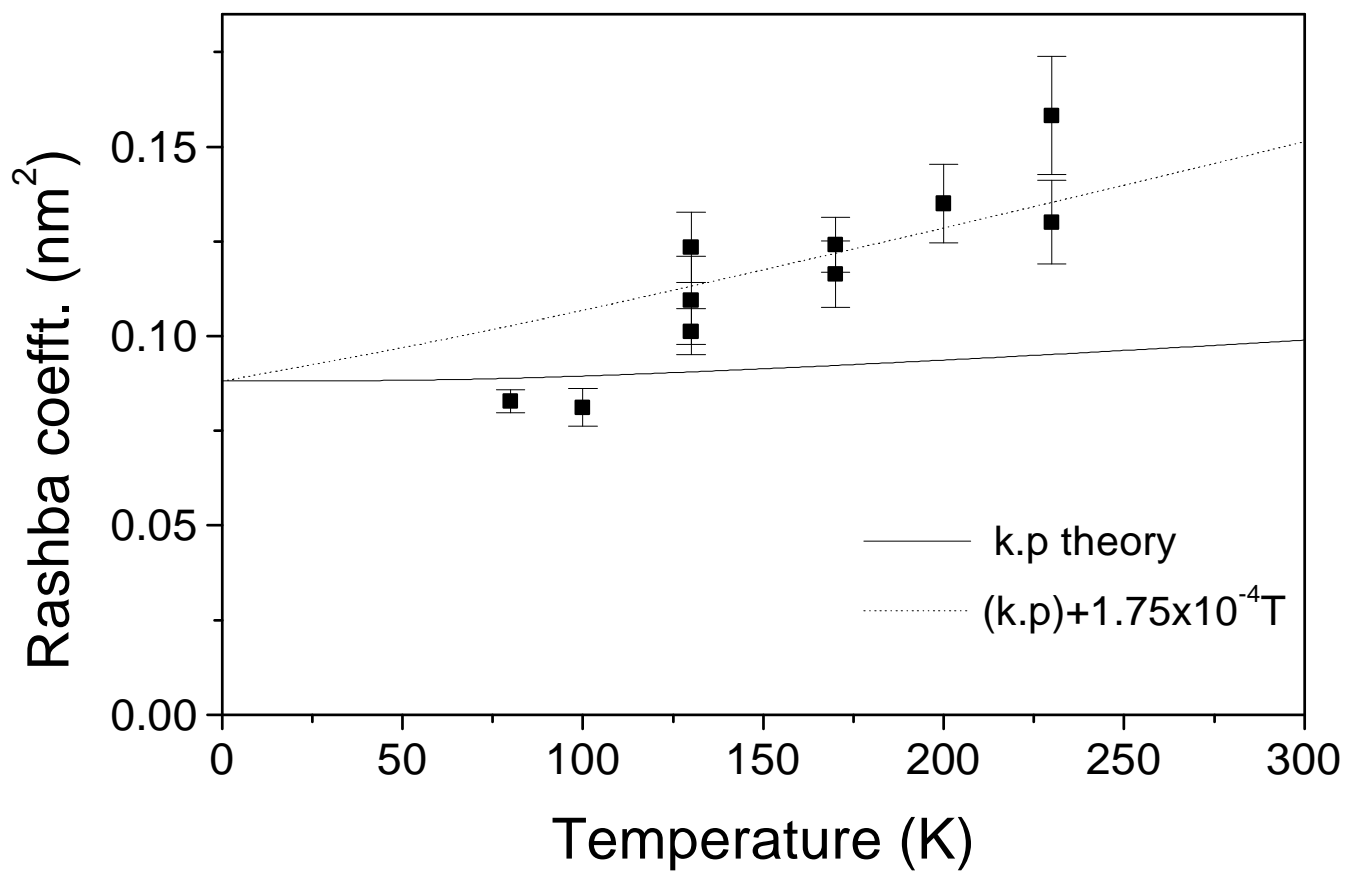

**Figure 4.** Rashba coefficient obtained from combination of spin relaxation and mobility measurements. Solid curve is 8-band **k.p** calculation including temperature dependence of band edges and of interband momentum matrix element. Dotted curve is k.p theory plus an empirical linear-in-$T$ term.